\documentclass[twocolumn,aps,amsmath,nofootinbib,superscriptaddress]{revtex4}

\usepackage{psfig}
\usepackage{graphicx}

\newcommand{\nn}{\nonumber}
\newcommand{\be}{\begin{equation}}
\newcommand{\bel}[1]{\begin{equation} \label{#1} }
\newcommand{\ee}{\end{equation}}
\newcommand{\bea}{\begin{eqnarray}}
\newcommand{\beal}[1]{\begin{eqnarray} \label{#1} }
\newcommand{\eea}{\end{eqnarray}}

\usepackage{color}

\definecolor{joerg}{rgb}{1.0,0.0,0.0}

\begin{document}

\title{Waking the Colored Plasma}

\author{J. Ruppert}
\email{ruppert@phy.duke.edu}
\affiliation{Department of Physics, Duke University, 
             Durham, NC 27708-0305, USA}

\author{B. M\"uller}
\email{muller@phy.duke.edu} 
\affiliation{Department of Physics, Duke University, 
             Durham, NC 27708-0305, USA}

\begin{abstract}
We calculate the wake induced in a hot, dense QCD medium by a fast parton in the framework of linear response theory. We discuss two different scenarios: a weakly coupled quark gluon plasma (pQGP) described by hard-thermal loop (HTL) perturbation theory and a strongly coupled QGP (sQGP), which had the properties of a quantum liquid.
\end{abstract}

\maketitle

\section{Introduction}
The quenching of QCD jets created in relativistic heavy ion collisions has been proposed as an important indicator for the creation of a quark-gluon plasma \cite{Gyulassy:1990ye,Wang:1991vs}. It is extensively studied theoretically and experimentally at RHIC.

The main emphasis in the theoretical studies has been on the description of the radiative energy loss which the leading parton suffers when traveling through the medium due to the emission of a secondary partonic shower. For an overview of the quantum field theoretical description of this process and its theoretical and experimental implications, see e.~g.~\cite{Accardi:2003gp,Jacobs:2004qv,Kovner:2003zj}. 

In this work we focus on another aspect of the in-medium jet physics, namely the fact that a charged particle with high momentum traveling through a medium induces a wake of current density, charge density and (chromo-)electric and magnetic field configurations. This wake reflects significant properties of the medium's response to the jet. 

We apply methods of linear response theory to the system of a relativistic color charge traveling through a QCD plasma. To investigate the coherence behavior of the plasma reacting to a charged particle traveling through the plasma, we consider the plasma's reponse to that external current. We restrict ourselves to only one weak, external current which points in a fixed direction in color group space, so that the field strength is linear in the current. In this framework quantum and non-abelian effects are included indirectly via the dielectric functions, $\epsilon_L$ and $\epsilon_T$. Linear response theory implies that the deviations from equilibrium are small enough that they will not, in turn, modify the dielectric function. Furthermore, for simplicity, we assume a homogenous and isotropic extension of the medium and disregard any finite size effects.

We discuss two qualitatively different scenarios: In the first one, we assume the plasma to be in the high temperature regime, where the gluon self-energies can be described using the leading order of the high temperature expansion, $T\gg \omega,k$. In the second scenario we investigate what happens if the plasma behaves as a strongly coupled quark gluon plasma (sQGP) which can be described as a quantum liquid. Our consideration of the second scenario has been motivated by earlier studies of wakes induced by fast electrically charged projectiles in the electron plasma of metallic targets \cite{Schaefer1,Schaefer2}, as well as by recent work exploring the induction of a conical flow by a jet in a sQGP \cite{Shuryak}. This latter study invokes a hydrodynamical description of the energy deposited by a quenched jet in the medium and emphasizes the emergence of a Mach cone. The idea of determining the sound velocity of the expanding plasma from the emission pattern of the plasma particles traveling at an angle with respect to the jet axis has also been mentioned in \cite{Stoecker}.

\section{Plasma physics}

The formalism for the linear response of the plasma can be straightforwardly generalized from electromagnetic plasma physics (see e.~g.~\cite{Ichimaru}). In this formulation, a dielectric medium is characterized by the components of the  dielectric tensor $\epsilon_{ij}(\omega,k)$, which for an isotropic and homogenous medium can be decomposed into the longitudinal and transverse dielectric functions $\epsilon_L$ and $\epsilon_T$ via:
\begin{eqnarray}
\epsilon_{ij}=\epsilon_L {\cal P}_{L,ij} +\epsilon_T {\cal P}_{T,ij}.
\end{eqnarray}
Here $ {\cal P}_{L,ij}=k_i k_j/ k^2$ and ${\cal P}_T=1-{\cal P}_L$ are the longitudinal and transverse orthonormal projectors with respect to the momentum vector $\vec{k}$. One can relate the dielectric functions $\epsilon_L$ and $\epsilon_T$ to the self-energies $\Pi_L$ and $\Pi_T$ of the in-medium gluon via \cite{LeBellac}:
\bea
\epsilon_L(\omega,k)=1-\frac{\Pi_L(\omega,k)}{K^2}, \\
\epsilon_T(\omega,k)=1-\frac{\Pi_T(\omega,k)}{\omega^2}\,,
\eea
where $K^2=\omega^2-k^2$.

Using Maxwell's equation and the continuity equation in momentum space the total chromoelectric field $\vec{E}^a_{\rm tot}$ in the QCD plasma is related to the external current $\vec{j}_{\rm ext}$ via:
 \begin{eqnarray} \label{EtotTOJ}
 \left[\epsilon_L {\cal P}_{L} + \left(\epsilon_T -\frac{k^2}{\omega^2}\right) {\cal P}_{T}\right] \vec{E}^a_{\rm tot}(\omega,\vec{k})=\frac{4 \pi}{i \omega} \vec{j}^a_{\rm ext} (\omega,\vec{k}) .
 \end{eqnarray}
Equation (\ref{EtotTOJ}) has non-trivial solutions only, if the determinant constructed from the elements of the tensor vanishes:
\bea
{\rm det}\left|\epsilon_L {\cal P}_{L} + (\epsilon_T -\frac{k^2}{\omega^2}) {\cal P}_{T}\right|=0 .
\eea
This equation governs the dispersion relation for the waves in the medium. Since it can be diagonalized into purely longitudinal and transverse parts, dispersion relations for the longitudinal and transverse dieelectric functions can be derived \cite{Ichimaru}:
\begin{subequations}
\bea \label{Dispersion}
\epsilon_L&=&0, \\
\epsilon_T&=&(k/\omega)^2.
\eea
\end{subequations}
These equations determine the longitudinal and transverse plasma modes. The longitudinal equation is also the dispersion relation for density-fluctuations in the plasma, namely space-charge fields which could be spontaneously excited in the plasma without an application of external disturbances \cite{Ichimaru}. 

The color charge density induced in the wake by the external charge distribution is:
\bea \label{charge}
\rho_{\rm ind}=\left(\frac{1}{\epsilon_L}-1\right)\rho_{\rm ext}.
\eea

The induced color charge density can also be calculated from the induced scalar potential via a Poisson equation:
\bea
\Phi_{\rm ind}= \frac{4\pi}{k^2} \rho_{\rm ind} ,
\eea
if one works in a gauge where the vector potential is transverse to the momentum \cite{Neufeld,Ichimaru}.

Since one can relate the total chromo-electric field to the induced charge in linear response theory by 
\bea
\vec{j}^a_{\rm ind}=i\omega(1-\epsilon)\vec{E}^a_{\rm tot}/(4\pi) ,
\eea
a direct relation between the external and the induced current can be derived using Eqn. (\ref{EtotTOJ}):
\bea \label{current}
\vec{j}^a_{\rm ind}=\left[\left(\frac{1}{\epsilon_L}-1\right){\cal P}_L + \frac{1-\epsilon_T}{\epsilon_T-\frac{k^2}{\omega^2}} {\cal P}_T\right]\vec{j}^a_{\rm ext}.
\eea
The induced charge and the induced current obey a continuity equation:
\bea \label{continuity}
i\vec{k}\cdot\vec{j}_{\rm ind}-i\omega \rho_{\rm ind}=0.
\eea

At this point we specify the current and charge densities associated with a color charge as Fourier transform of a point charge moving along a straight-line trajectory with constant velocity $\vec v$:
\begin{subequations}
\begin{eqnarray}
\vec{j}^a_{\rm ext}&=&2\pi q^a \vec{v} \delta(\omega-\vec{v} \cdot \vec{k}),\\
\vec{\rho}^a_{\rm ext}&=&2\pi q^a \delta(\omega-\vec{v} \cdot \vec{k})
\end{eqnarray}
\end{subequations}
 where $q^a$ its color charge defined by $q^a q^a = C \alpha_s$ with the strong coupling constant $\alpha_s=g^2/4\pi$ and the quadratic Casimir invariant $C$ (which is either $C_F=4/3$ for quarks or $C_A=3$ for a gluon). In this simplified model one disregards changes of the color charge while the particle is propagating through the medium by fixing the charge's orientation in color space \cite{Weldon,Thoma}.
 
 The non-radiative part of the energy loss of the incident color charge is given by the back reaction of the induced chromo-electric field onto the incident parton. The energy loss per unit length is given by \cite{Ichimaru, Thoma}:
\bea \label{energy1}
\frac{dE}{dx}=q^a \frac{\vec{v}}{v} {\rm Re} \vec{E}^a_{\rm ind}(\vec{x}=\vec{v}t,t),
\eea
where the induced chromo-electric field $\vec{E}^a_{\rm ind}$ is the total chromo-electic field minus the vacuum contribution. Using the inverse of Eqn. (\ref{EtotTOJ}) it is given by:
\bea \label{loss}
\vec{E}^a_{\rm ind}=&&\left[\left(\frac{1}{\epsilon_L}-1\right) {\cal P}_L + \right. \nn \\ && \left. \left(\frac{1}{\epsilon_T-\frac{k^2}{\omega^2}}-\frac{1}{1-\frac{k^2}{\omega^2} }\right)  {\cal P}_T\right] \frac{4\pi}{i\omega}\vec{j}^a_{\rm ext}.
\nn \\ 
\eea
From (\ref{energy1}) and ({\ref{loss}) the non-radiative energy loss per unit length is given by \cite{Thoma}:
\bea \label{energy2}
\frac{dE}{dx}=-\frac{C \alpha_s}{2 \pi^2 v} \int d^3k \frac{\omega}{k^2}({\rm Im} \epsilon_L^{-1}+  \nn \\ (v^2 k^2 - \omega^2){\rm Im} (\omega^2 \epsilon_T^{-1}- k^2)^{-1}),
\eea where $\omega=\vec{v}\cdot\vec{k}$.

\section{Charge wake in a QGP in the high temperature approximation}
 
In this section we study the first scenario in detail. The plasma is assumed to be in the high temperature regime, where the gluon self-energies can be described by the so-called hard-thermal loop (HTL) approximation, the leading order of the high temperature expansion $T\gg \omega, k$ \cite{Klimov, Weldon}. This regime can be expected to be realized above the deconfinement temperature $T_c$. The self-energies derived in this approximation have been shown to be gauge invariant \cite{Pisarski} and the dielectric functions obtained from these are therefore also gauge invariant. 
The dielectric functions read explicitly:
\bea
 \epsilon_L &=& 
 1+\frac{2m_g^2}{k^2}
 \left[
 1-\frac{1}{2}x
 \left(
 {\rm ln}\left|\frac{x+1}{x-1}\right|-i\pi \Theta \left(1-x^2\right)
 \right)
 \right], 
 \nn \\
 \epsilon_T&=&1-\frac{m_g^2}{\omega^2} 
 \left[ 
 x^2+
 \frac{x(1-x^2)}{2} \times \right.  \nn \\ && 
 \left.
 \left(
 {\rm ln}\left|\frac{x+1}{x-1}\right|-i\pi 
 \Theta 
 \left(
 1-x^2 
 \right) 
 \right)
  \right],
\label{eps}
\eea

where $x=\omega/k$. 
We first discuss the induced charge and current densities $\rho_{\rm ind}$ and $\vec{j}_{\rm ind}$ using eqs.~(\ref{charge},\ref{current}). 
The Fourier transform of (\ref{charge}) reads in cylindrical coordinates:
\bea \label{charge2}
\rho_{\rm v, ind}(\rho, z, t)=\frac{m_g^3}{(2 \pi)^2 v} q^a \int_0^\infty d\kappa' \kappa' J_0(\kappa' \rho m_g) \times \nn \\ \int^{\infty}_{-\infty} d\omega' \,{\rm exp} \left[i\omega'\left(\frac{z}{v}-t\right)m_g\right] \left(\frac{1}{\epsilon_{L}}-1 \right),
\eea
 where $k=\sqrt{\kappa^2+\omega^2/v^2}$, $\kappa=\kappa' m_g$ and $\omega=\omega' m_g$, showing that the induced charge density $\rho_{\rm v, ind}$ is proportional to $m_g^3$. The cylindrical symmetry around the jet axis restricts the form of the current density vector $\vec{j}_{\rm ind}$. It has only non-vanishing components parallel to the beam axis, $\vec{j}_{{\rm v}, {\rm ind}}$, and radially perpendicular to it, $\vec{j}_{\rho, {\rm ind}}$:
\begin{widetext}
\bea \label{current2}
\vec{j}_{\rm v, ind}(\rho, z,t) &=&
\frac{m_g^3}{(2 \pi)^2 v^2} q^a  \int_0^\infty d\kappa' \kappa' J_0(\kappa' \rho m_g) \int^{\infty}_{-\infty} d\omega' {\rm exp} \left[i\omega'\left(\frac{z}{v}-t\right)m_g\right] \left[\left(\frac{1}{\epsilon_{L}}-1\right) \frac{\omega^2}{k^2} + \frac{1-\epsilon_{T}}{\epsilon_{T}-\frac{k^2}{\omega^2}}\left(v^2-\frac{\omega^2}{k^2}\right) \right], 
\nn \\
\vec{j}_{\rho, ind}(\rho, z ,t) &=& 
\frac{i m_g^3}{(2 \pi)^2 v} q^a \int_0^\infty d\kappa' \kappa' J_1(\kappa' \rho m_g) \int^{\infty}_{-\infty} d\omega' {\rm exp} \left[i\omega'\left(\frac{z}{v}-t\right)m_g\right] \frac{\omega \kappa}{k^2}\left[\left(\frac{1}{\epsilon_{L}}-1\right)-\left(\frac{1-\epsilon_{T}}{\epsilon_{T}-\frac{k^2}{\omega^2}}\right)\right] .
\eea
\end{widetext}
Again, the components of the current density are proportional to $m_g^3$. These expressions are only valid for $v>0$. For a parton at rest the current density vector vanishes and the induced charge density can be directly derived from Eq. (\ref{charge}). It has a Yukawa-like shape and reads (for $r=|\vec{r}|$): 
\bea 
\rho_{\rm v=0, ind}(r)=- \frac{m_g^2}{2 \pi r} {\rm exp}(-\sqrt{2}m_g r) q^a.
\eea

Since longitudinal and transverse plasma modes, as determined by the dispersion relations (\ref{eps}), can only appear in the time-like sector of the $\omega,k$ plane \cite{Weldon, LeBellac}, collective excitations do not contribute to the charge and current density profile of the wake. Emission analogous to Cherenkov radiation and Mach cones do not appear, but the charge carries a screening color cloud along with it. Fig.~\ref{figure1} illustrates this physically intuitive result numerically.  It shows the charge density of a colored parton traveling with $v=0.99 c$ in cylindrical coordinates.

\begin{figure}
 \par\resizebox*{!}{0.50\textheight}{\includegraphics{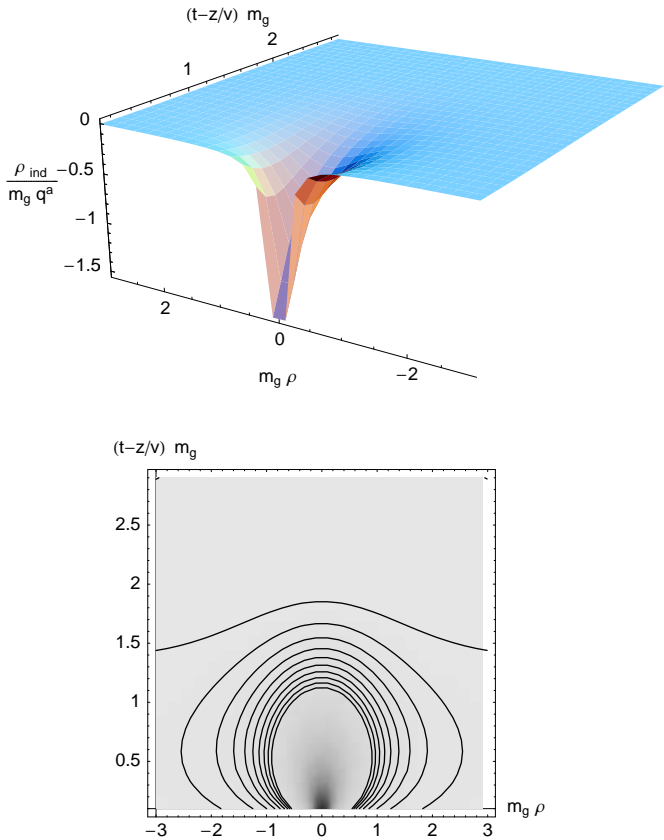}} \par{}
\caption{Spatial distribution of the induced charge density from a jet with fixed color charge $q^a$ in a high temperature plasma where the HTL approximation is applicable. The lower plot shows equi-charge lines in the density distribution. The density profile is a cloud traveling with the color charged jet. 
\label{figure1}}
\end{figure}
 
Despite the fact that Mach cones do not appear in the charge density wake,
the particle still suffers energy loss due to elastic collisions in the medium, which can be described by eq.~(\ref{energy2}). This effect of energy dissipation has been studied in \cite{Thoma} and we refer the reader to this work for details. The integrand in (\ref{energy2}) contributes to the integral in the space-like region only, where $|x|<1$, and therefore does not get contributions from frequencies where collective plasma modes exist. This is consistent with the fact that such modes are not excited in the induced charge and current densities.

\section{Charge wake induced in a strongly coupled QGP}

In this section we consider another scenario and assume that the plasma is in a strongly coupled regime where it exhibits properties characteristic of a quantum liquid. We investigate the similarities and differences in the wake of a colored parton in such a quantum liquid with those obtained for the perturbative, ultrarelativistic plasma discussed in the last section. Since up to date there is a lack of theoretical methods for first principle calculations of the color response functions in a strong coupled QGP regime, we restrict our discussion in this section to a simple model. Nevertheless, we construct this simplified model in such a way that important general conclusions about the differences and similarities of a quantum liquid scenario and the scenario of the last section can be drawn. 

The most prominent difference - in accordance with a sQGP quantum liquid scenario - is the possibility of a plasmon mode that extends into the space-like region of the $\omega,  k$ plane above some $k$.

The sQGP paradigm suggests very low dissipation at small $k$ ("hydro modes"), but large dissipation at high $k$. Our assumption is that a critical momentum $k_c$ separates the regimes of collective and single particle excitation modes in the quantum liquid where the dominant modes below $k_c$ are plasmon excitations and where dissipative single-particle response is negligible. Since we are predominantly interested in collective effects in the plasma here, we restrict our study to the region below $k_c$.

To be specific, we assume that the dielectric function of the strongly coupled plasma in the $k<k_c$ regime leads to a longitudinal dispersion relation of the form: 
\bea \label{Dispersion2}
\omega_{\rm L}=\sqrt{u^2 k^2 + \omega_p^2}\,\,,
\eea
where $\omega_p$ denotes the plasma frequency and $u$ the speed of plasmon propagation which is assumed to be constant. In accordance with (\ref{Dispersion2}) we assume the following dielectric function
\bea \label{NonBloch}
\epsilon_L=1+\frac{\omega_p^2/2}{u^2 k^2 - \omega^2 + \omega_p^2/2}\,\, \, \, (k\le k_c)  \, \,.
\eea Note that this differs from the classical, hydrodynamical dielectric function of Bloch \cite{Bloch}, since the latter one is singular at small $k$ and $\omega$ due to phonon contributions which cannot mix with colored plasmons.
The dielectric function (\ref{NonBloch})  is constructed in such a way that it allows to study one possible aspect of a quantum liquid scenario, namely a plasmon mode which extends into the space-like region of the $\omega, k$ plane. The qualitative induced wake structure in such a quantum liquid scenario is general, namely a Mach shock wave structure for a supersonically traveling color source. In that respect the principlal findings of this and the next section can be expected to hold generally for a quantum liquid with a plasmon branch similiar to (\ref{NonBloch}) independent of the exact form of the dielectric function.

We assume a speed of plasmon propagation of $u/c= 1/\sqrt{3}$. This differs from the speed of plasmon propagation obtained in the HTL approximation in the small $k$ limit \cite{Weldon}, namely $u/c=\sqrt{3/5}$, and of a hadronic resonance gas $u/c \approx \sqrt{0.2}$ \cite{Shuryak2, Venugopalan}. We disregard possible damping effects for $k<k_c$. This assumption implies a negligible imaginary part in the dielectric function of the strongly coupled QGP, i.~e.\ no collisional energy loss in the collective regime.

Determining the plasmon mode via Eq.~(\ref{Dispersion2}) reveals that in this scenario the mode is in the space-like region of the $\omega, k$ plane for $k>\omega_p/(\sqrt{1-u^2})$ . Recall that this is different for the high-temperature plasma, where longitudinal and transverse plasma modes only appear in the time-like region, $|x|=|\omega/k|>1$.  In the quantum liquid scenario one can expect that the modes with low phase velocity $|x|<u$ suffer severe Landau damping because they accelerate the slower moving charges and decelerate those moving faster than the wave \cite{Weldon, Ichimaru}. A charge moving with a velocity that is lower than the speed of plasmon propagation can only excite these modes and not the modes with intermediate phase velocities $u<|x|<1$, which are undamped \cite{Ichimaru, Weldon}. The qualitative properties of the color wake are in this case analogous to those of the high temperature plasma, namely that the charge carries only a screening color cloud with it and Cherenkov emission and Mach cones do not appear. Using eq.~(\ref{charge2}) and restricting the integration area to the region $k<k_c=2 \omega_p$,\footnote{Note that the expressions in \cite{Schaefer1, Schaefer2} do not correspond to $k<k_c$, but $\kappa<k_c$.} one can illustrate this for a colored particle traveling with $v=0.55 c<u$. Fig.~\ref{figure2} shows the charge density cloud traveling with the colored parton.

\begin{figure}
 \par\resizebox*{!}{0.38\textheight}{\includegraphics{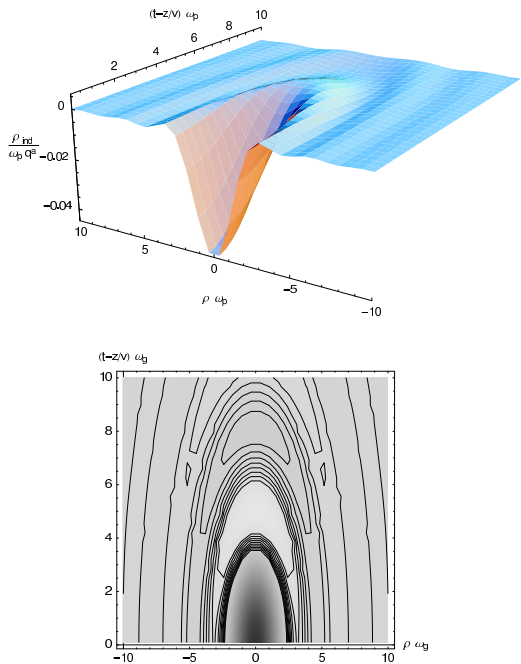}} \par{}
\caption{Spatial distribution of a induced charge density from a  jet with 
high momentum and a fixed color charge $q^a$ that is traveling with $v=0.55c<u$. The lower plot shows equi-charge lines in the density distribution. The density profile is that of a cloud traveling with the color charged jet. 
\label{figure2}}
\end{figure}

If the colored parton travels with a velocity $v>u$ that is higher than the speed of sound in the medium, modes with an intermediate phase velocity $u<|x|<1$ can be excited. The emission of these plasma oscillations induced by supersonically traveling particles is analogous to Cherenkov radiation. This leads to the emergence of Mach cones in the induced charge density cloud with the opening angle
\bea \label{Mach}
\Delta \Phi={\rm arccos}\left(u/v\right)
\eea

The effect is well known in solid state physics \cite{Ichimaru} and is analogous to the Mach cones induced by fast heavy ions in electron plasmas \cite{Schaefer1, Schaefer2, Groeneveld}. The physics of shock waves in relativistic heavy-ion collisions has also been discussed in \cite{Glassgold, Scheid:1973, Baumgardt:1975qv}.
 
Figures \ref{figure31}a,b show the induced charge density for a colored particle traveling at $v=0.99 c>u$. Figs.~\ref{figure32}a,b show the corresponding current densities. Here eqs.~(\ref{charge2},\ref{current2}) have been used, and the integration area has been restricted to the region $k<k_c=2 \omega_p$. Numerical consistency has been checked also via the continuity equation (\ref{continuity}).

\begin{figure}
 \par\resizebox*{!}{0.5\textheight}{\includegraphics{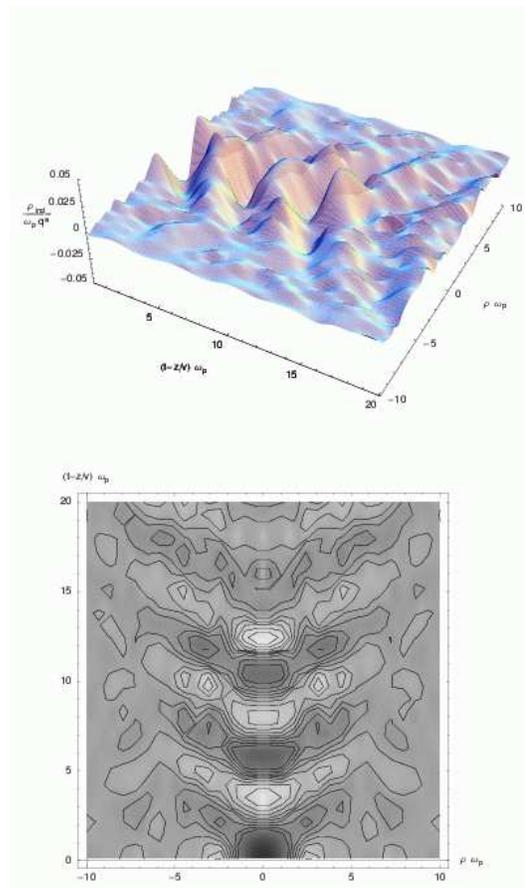}} \par{}
\caption{(a) Spatial distribution of the induced charge density from a  jet with high momentum and fixed color charge $q^a$ that is traveling with 
$v=0.99c>u=\sqrt{1/3}c$. (b) Plot showing equi-charge lines in the density distribution for the situation in (a). 
\label{figure31}}
\end{figure}

\begin{figure}
 \par\resizebox*{!}{0.5\textheight}{\includegraphics{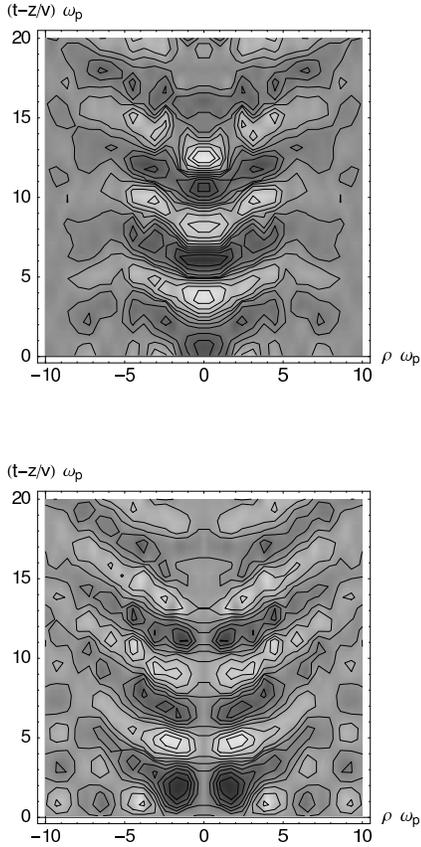}} \par{}
\caption{(a) Spatial distribution of the equi-value lines of the current densitie's component parallel to a jet travelling with $v=0.99c>u=\sqrt{1/3}c$ in a sQGP.
 (b) Spatial distribution of the equi-value lines of the current densitie's component perpendicular to a jet travelling with $v=0.99c>u=\sqrt{1/3}c$ in a sQGP.
\label{figure32}}
\end{figure}

We emphasize that the existence of Mach cones is expected in a plasma in general if the particle is moving faster than the speed of sound in the plasma and if the dispersion relation of the collective mode extends into the space-like region. The wake induced by a colored jet in such a setting leads to regions of enhanced and depleted charge density in the wake, which have the shape of Mach waves trailing the projectile.

\section{Observable consequences}

It can be expected that the phenomenon of Mach cones should eventually lead to a directed emission of secondary partons from the plasma. This has its anology in solid state physics where Mach cones induced by fast heavy ions in an electron plasma lead to an emission spectrum of electrons carried within the wake \cite{Schaefer1, Schaefer2} which have been studied experimentally \cite{Groeneveld}.

If that scenario of a strongly coupled QCD plasma is realized in the matter created in an ultrarelativistic heavy ion collision, one could expect to observe these cones in the actual distribution of secondary particles associated with jets at RHIC, a signature also proposed in \cite{Stoecker, Shuryak}. 

Indeed preliminary data from the STAR experiment (see Fig.~1 in \cite{STAR}) (also PHENIX data can be expected to be available regarding this issue) show such effects in the background distribution of secondary particles in the azimuthal angle $\Delta \phi$. The peak near zero degree corresponds to secondaries from an outmoving jet, particles from the companion jet result in a distribution with a clear maximum near $\Delta \phi=\pi$ for pp collisions, where no medium effects are present. In Au-Au collisions there is a distribution with two maxima around the $\Delta \phi=\pi$ position. These are located at $\Delta \phi \approx \pi  \pm 1.1$.  One can argue \cite{Shuryak} that such an effect could possibly be traced back to a Mach shock front traveling with the side-away jet  leading to maxima in the distribution at about  $\Delta \Phi = \pi \pm {\rm arccos} (u/v)$. 
Given the confirmation of this effect in the data, this would clearly indicate that the first scenario, viz.~a plasma described by perturbative color dielectric functions, is not realized in the RHIC experiments. Furthermore, constraints of the properties of the plasmons in general could be deduced. 

The preliminary data indicate that a pronounced Mach cone might be reflected only in the secondaries of the {\it quenched} jet. This can probably be attributed to the circumstance, that the unquenched jets are emitted points too close to the plasma surface to generate a fully formed wake behind them. 

The speed of plasma propagation could be also determined. In fact if the maxima at $\Delta \phi \approx \pi \pm 1.1$ are experimentally confirmed, it would correspond to $u/c\approx\sqrt{0.2}$ \cite{Shuryak}. It is interesting to note, that a study of  the angular structure of the collisional energy losses of a hard jet in the medium would also support such an observation: the incident hard jet's scattering angle vanishes in the relativistic limit - leading to the jet's propagation along a straight line - whereas the expectation value of the scattering angle $\Theta$ of a struck "thermal" particle is $\left<\Theta\right>\approx 1.04$ \cite{Lokhtin} which is close to $1.1$.

\section{Conclusions}

In this letter we discussed the properties of the charge density wake of a colored hard partonic jet traveling  through a QGP plasma in the framework of linear response theory. We have studied two different scenarios, namely high temperature QGP at $T \gg T_c$ described in the HTL approximation, and a description of a strongly coupled QGP  (sQGP) behaving as a quantum liquid. We found that the structure of the wake corresponds to a screening color cloud traveling with the particle in the case of the high temperature plasma and in the case of a quantum fluid, if the velocity of sound in the plasma is not exceeded by the jet in the latter case. The structure of the wake is changed considerably in comparison to the former cases, if the jet's velocity exceeds the plasma's speed of sound and the collective modes have a dispersion relation extending in the space-like region. Then the induced parton density exhibits the characteristics of Mach waves trailing the jet at the Mach angle. 

It is argued that this effect could be used to constrain theoretically possible scenarios by experimental analysis via the angular distribution of the secondary particle cones of jet events in RHIC. First indications of the observation of a Mach shock phenomenon in RHIC in the quenched jet's secondary particle distribution from STAR data were discussed.

In general secondary particle distributions can be used to provide methods of probing the QCD plasma's collective excitations experimentally.
\bigskip

\section*{Acknowledgements}
This work was supported in part by U.~S.~Department of Energy under grants DE-FG02-96ER40945 and DE-FG02-05ER41367. JR thanks the Alexander von Humboldt Foundation for support as a Feodor Lynen Fellow. 

\bibliography{u4}

\end{document}